\input ./style/arxiv-general.cfg
\documentclass[aoas,MSNbibl,nameyear,seceqn,dvips]{arximspdf}
\makeatletter
   \@ifpackageloaded{graphicx}{}{\usepackage{graphicx}}
\makeatother
\usepackage{dcolumn}

\doi{10.1214/15-AOAS865}
\volume{9}
\issue{4}
\pubyear{2015}
\firstpage{2052}
\lastpage{2072}
\docsubty{FLA}

\makeatletter
\newcolumntype{d}[1]{D{.}{.}{#1}}
\newcommand{\eqref}[1]{(\ref{#1})}
\makeatother

\begin{document}
\begin{frontmatter}

\title{Identifying heterogeneous transgenerational DNA methylation
sites via clustering in beta regression\thanksref{T1}}
\runtitle{Identifying transgenerational methylation sites}

\begin{aug}
\author[A]{\fnms{Shengtong}~\snm{Han}\thanksref{M1,T4}\ead[label=e1]{shengtonghan@gmail.com}},
\author[A]{\fnms{Hongmei}~\snm{Zhang}\corref{}\thanksref{M1,T2}\ead[label=e21]{hzhang6@memphis.edu}},\break
\author[B]{\fnms{Gabrielle~A.}~\snm{Lockett}\thanksref{M2,T4}\ead[label=e31]{G.A.Lockett@soton.ac.uk}},
\author[A]{\fnms{Nandini}~\snm{Mukherjee}\thanksref{M1,T4}\ead[label=e32]{nmkhrjee@memphis.edu}},
\author[B]{\fnms{John~W.}~\snm{Holloway}\thanksref{M2,T4}\ead[label=e33]{J.W.Holloway@soton.ac.uk}} 
\and
\author[A]{\fnms{Wilfried}~\snm{Karmaus}\thanksref{M1,T3}\ead[label=e22]{karmaus1@memphis.edu}}
\runauthor{S. Han et al.}
\affiliation{University of Memphis\thanksmark{M1} and
University of Southampton \thanksmark{M2}}
\address[A]{S. Han\\
H. Zhang\\
W. Karmaus\\
N. Mukherjee\\
Division of Epidemiology, Biostatistics,\\
\quad and Environmental Health\\
School of Public Health\\
University of Memphis\\
Memphis, Tennessee 38111\\
USA\\
\printead{e1}\\
\phantom{E-mail: } \printead*{e21}\\
\phantom{E-mail: } \printead*{e22}\\
\phantom{E-mail: } \printead*{e32}}
\address[B]{G. A. Lockett\\
Human Development and Health\\
Faculty of Medicine\\
University of Southampton\\
Southampton SO16 6YD\\
United Kingdom\\
\printead{e31}}
\address[C]{J. W. Holloway\\
Human Development and Health\\
and\\
Clinical and Experimental Sciences\\
Faculty of Medicine\\
University of Southampton\\
Southampton SO16 6YD\\
United Kingdom\\
\printead{e33}}
\end{aug}
\thankstext{T1}{Supported by National Institutes of Health.}
\thankstext{T4}{Supported by NIH R01AI091905.}
\thankstext{T2}{Supported by NIH R21AI099367 and NIH R01AI091905.}
\thankstext{T3}{Supported by NIH R01AI091905.}

%
\received{\smonth{10} \syear{2014}}
%
\revised{\smonth{8} \syear{2015}}

%
\begin{abstract}
This paper explores the transgenerational DNA methylation pattern (DNA
methylation transmitted
from one generation to the next) via a clustering approach. Beta
regression is employed to model the transmission pattern from parents
to their offsprings at the population level. To facilitate this goal,
an expectation maximization algorithm for parameter estimation along
with a BIC criterion to determine the number of clusters is proposed.
Applying our method to the DNA methylation data composed of 4063~CpG
sites of 41 mother--father-infant triads, we identified a set of~CpG
sites in which DNA methylation transmission is dominated by fathers,
while at a large number of~CpG sites, DNA methylation is mainly
maternally transmitted to the offspring.
\end{abstract}

%
\end{frontmatter}

\setcounter{footnote}{4}

\section{Introduction}\label{Intro}

Genetics strongly influences allergic disease risk, yet loci identified
throughout genome-wide association studies (GWAS) cannot fully explain
disease heritability, a phenomenon known as ``missing heritability''
\citep{TeriA.Manolio2009461}. Transgenerational transmission of
epigenetic marks (epigenetics transmitted from one generation to the
next) such as DNA methylation is one possible mechanism accounting for
the missing heritability \citep{GabrielleALockett20135}. DNA
methylation, that occurs at Cytosine--Guanine (CpG) dinucleotides, has
been strongly associated with health outcomes, including allergic
diseases such as eczema, asthma and rhinitis \citep
{ColmE.Nestor201410,Yousefi20134,NelisSoto-Ramirez20135,Zhang20146,A.H.Ziyab201227}.
Furthermore, the observation of asymmetric transmission from parent to
child of allergic diseases \citep{S.HasanArshad2012130} suggests the
potential involvement of epigenetic mechanisms in the inheritance of
allergic disease.

There is evidence that DNA methylation patterns can be transmitted to
the next generation (i.e., transgenerational transmission) in
mammals: a recent study in mice found that maternal folate restriction
produces congenital malformations which persist into the fifth
generation after exposure, likely through epigenetic mechanisms \citep
{NishaPadmanabhan2013155}. Research findings also indicate that
transgenerational transmission of famine responses in humans is
believed to be mediated by epigenetic mechanisms \citep
{MarcusEPembrey200614,GunnarKaati200715}. However, we know very little
about the transmission mechanisms, that is, which~CpG sites display
transmission of DNA methylation to the next generation dominated by
inheritance of methylation from the mother, and which~CpG sites'
inheritance dominated by father (asymmetric transmission). Uncovering
the transmission pattern of DNA methylation from parents to offspring
in the general population and identifying~CpG sites whose DNA
methylation is asymmetrically transmitted will provide the potential
for allergic disease prediction as well as prevention \citep
{Lockett20135,Szyf200949}.

In this study, we tackle this problem by evaluating the transmission
pattern via Beta regressions and grouping~CpG sites if they share a
similar pattern of DNA methylation transmission from parents to their
offspring. The grouping is fulfilled via clustering of~CpG sites based
on the association of offspring's mean DNA methylation with parents'
mean DNA methylation at the population level using a Beta regression
\citep{SilviaFerrari200431} (i.e., the association is at the population
level). We not only identify~CpG sites showing heterogeneous
transmission patterns but also identify~CpG sites showing similar
transmission patterns at the population level. Traditional model-based
methods for clustering variables are built into normal distributions
and focus on associations in individual subjects (i.e., at the
individual level) \citep{Qin200662}. On the other hand, unsupervised
methods, such as the $K$-means algorithm \citep
{MACQUEEN19671,Hartigan197928}, partitioning around medoids \citep
{Park200936} or various hierarchical clustering methods, are not able
to evaluate the strength of inheritance while clustering. To the best
of our knowledge, very limited contribution has been made to this field.

In this article, methodology, including model assumption and the
expectation and maximization (EM) algorithm, is presented in
Section~\ref{Meth}. Section~\ref{Nume} discusses simulations and real
data applications, where we compared our method via simulations with
the commonly used $K$-means approach. Summary and discussions are given
in Section~\ref{SuDi}.

\section{Methodology}\label{Meth}
\subsection{Model assumption}\label{modelaas}
Suppose there are $I$ triads (each triad consists of one child and the
child's two parents) and $J$~CpG sites which are common to all triads.
We further assume that all~CpG sites are independent of each other. Let
$Z1_{ij}$ and $Z2_{ij}$ denote DNA methylation at~CpG site $j$ for the
$i$th offspring's mother and father ($F_1$ generation) with the domain
on interval $(0, 1)$, respectively, which could be assumed to follow
Beta distributions \citep{Houseman20089}
\[
Z1_{ij}\sim \operatorname{Beta}\bigl(\alpha_j^M,
\beta_j^M\bigr),\qquad Z2_{ij}\sim \operatorname{Beta}\bigl(
\alpha_j^F, \beta_j^F\bigr),
\]
where $0< Z1_{ij}, Z2_{ij}<1$, $\alpha_j^M, \beta_j^M, \alpha_j^F, \beta
_j^F$ are\vspace*{2pt} unknown scale parameters, $i=1, \ldots, I; j=1, \ldots, J $.
Let $y_{ij}$ denote the methylation score at site $j$ of child $i$
($F_2$ generation). Conditional on the DNA methylation in parents due
to inheritance, $y_{ij}$ satisfies
\[
y_{ij}|Z1_{ij},Z2_{ij} \sim \operatorname{Beta}\bigl(
\alpha_j^0, \beta_j^0\bigr),
\]
where\vspace*{1pt} $0<y_{ij} <1$, $\alpha_j^0$ and $\beta_j^0$ are two unknown scale
parameters in the Beta distribution.\vspace*{1pt} Let $O_j=\operatorname{logit}(\frac{\alpha
_j^0}{\alpha_j^0+\beta_j^0})=\log(\alpha_j^0)-\log(\beta_j^0)$,
$M_j=\log(\alpha_j^M)-\log(\beta_j^M)$ and $F_j=\log(\alpha_j^F)-\log(\beta
_j^F)$ denote logit transformed mean methylation of site $j$ for child,
mother and father, respectively. The inheritance relation from parents
to their offspring in a general population is assumed to be
%
\begin{equation}
O_j=\gamma_{0j}+\gamma_{1j}M_j+
\gamma_{2j}F_j, \label{inheritance}
\end{equation}
where $\gamma_{0j}$ is the intercept, and $\gamma_{1j}, \gamma_{2j}$
represent the inheritance strength from mother and father to offspring,
respectively. It is worth pointing out that this dependence is for each
individual~CpG site. The assumption of independence between~CpG sites
noted earlier is not likely to influence the dependence structure
between parents and offspring at each individual site. Note that in
(\ref{inheritance}) we assumed a linear association as well as additive
parental effects. The linearity assumption is supported by our
preliminary data in terms of correlations (as seen in our real data
application) and epigenetic inheritance studies \citep{Rakyan2003100}.
Model (\ref{inheritance}) describes a manifestation of parental DNA
methylation inheritance to offspring. It is possible that the effects
may be multiplicative or in another unknown format, which certainly
deserves further investigation, and we hope our attempt is a starting
point of this exploration. It is not unusual that some~CpG sites share
the same transmission pattern in terms of $\gamma_{1j}, \gamma_{2j}$.
Identifying~CpG sites following similar transmission patterns will
improve our understanding of related genes or biological pathways
involved in DNA methylation transmission. To this end, we perform
cluster analysis and revise \eqref{inheritance} as
\[
O_{j}=\gamma_{0k}+\gamma_{1k}M_{j}+
\gamma_{2k}F_{j}, \label{relation}
\]
for~CpG site $j$ in cluster $k, k=1, 2, \ldots, K$. Denote by $\bolds
{\gamma}_k=(\gamma_{0k}, \gamma_{1k}, \gamma_{2k})$ the transmission
coefficients in cluster $k$. All~CpG sites in cluster $k$ share the
same transmission pattern from the $F_1$ generation to $F_2$
generation. If $\gamma_{1k}=0$ or $\gamma_{2k}=0$, then a child's DNA
methylation at these~CpG sites is inherited from his/her father alone
or mother alone. If $\gamma_{1k}=\gamma_{2k}=0$, there is no
transmission from parents and average DNA methylation of a child is
determined by $\gamma_{0k}$. In other situations, both parents transmit
their methylation to their child but possibly with different strengths.
It is reasonable to assume that the number of clusters $K$ is
substantially smaller than the number of~CpG sites $J$, $K \ll J$.
Selection of the number of clusters $K$ is presented in Section~\ref
{numclust}. To infer the parameters $\bolds{\gamma}_k$ and scale
parameters in Beta distributions, we introduce the following empirical
expectation maximization (EM) algorithm.

\subsection{The likelihood function and the empirical EM algorithm for
clustering}\label{emsection}
We start from introducing necessary notation.
Let $\bolds{\mu}=(\bolds{\mu}_1, \bolds{\mu}_2, \ldots, \bolds{\mu}_J)$
with $\bolds\mu_j=(\mu_{j1}, \mu_{j2}, \ldots, \mu_{jK})^T=(0, 0,
\ldots, 1, 0, \ldots, 0)^T$, a $K \times1$ vector and $\mu_{jk}=1$
indicating site $j$ is in cluster $k$. Denote by $\pi_k$ the
probability of each site falling into cluster $k$ and it is free of
site index $i$. We assume $\bolds{\mu}_i\sim \operatorname{Mult}(1, \bolds{\pi})$
(Multinomial distribution), where $\bolds{\pi}=(\pi_1, \pi_2, \ldots,
\pi_K)$, with $0 \leq\pi_k\leq1$, $k=1, 2, \ldots, K$, $\sum_{k=1}^K
\pi_k=1$. Let $\bolds{\theta}=(\bolds{\alpha}, \bolds{\beta}, \bolds
{\gamma}, \bolds{\pi})$ denote a collection of all parameters, where
$\bolds{\alpha}=(\bolds{\alpha}^0, \bolds{\alpha}^M, \bolds{\alpha}^F)$
with $\bolds{\alpha}^0=(\alpha_1^0, \alpha_2^0, \ldots, \alpha_J^0)$,
$\bolds{\alpha}^M=(\alpha_1^M, \alpha_2^M, \ldots, \alpha_J^M)$ and
$\bolds{\alpha}^F=(\alpha_1^F, \alpha_2^F, \ldots, \alpha_J^F)$.
Analogous to $\bolds{\alpha}$, parameter $\bolds{\beta}$ has the same
structure, $\bolds{\beta}=(
\bolds{\beta}^0,
\bolds{\beta}^M, \bolds{\beta}^F)$ with $\bolds{\beta}^0=(\beta_1^0,
\beta_2^0, \ldots, \beta_J^0)$, $\bolds{\beta}^M=(\beta_1^M, \beta_2^M,
\ldots, \beta_J^M)$ and $\bolds{\beta}^F=(\beta_1^F, \beta_2^F, \ldots,
\beta_J^F)$.
Finally, parameter $\bolds{\gamma}$ is a collection of coefficients,
$\bolds{\gamma}=(\bolds{\gamma}_1, \bolds{\gamma}_2, \ldots, \bolds
{\gamma}_K)$ with
$\bolds{\gamma}_k=(\gamma_{0k}, \gamma_{1k}, \gamma_{2k}), k=1, 2,
\ldots, K$. Denote by $\mathbf{Y}=(y_{ij}, Z1_{ij}, Z2_{ij}), i=1, 2,
\ldots, I, j=1, 2, \ldots, J$, the observed data. The likelihood of
$\bolds{\theta}$ is
\begin{eqnarray*}
\label{observedlikelihood} L(\bolds{\theta}|\mathbf{Y}) &=&\prod
_{i=1}^I \prod_{j=1}^J
\prod_{k=1}^K P(\mu_{jk}|\bolds
\theta)P(y_{ij}|Z1_{ij},Z2_{ij},\mu_{jk},
\bolds\theta)^{\mu_{jk}}P(Z1_{ij}|\bolds\theta)P(Z2_{ij}|\bolds
\theta)
\\[-2pt]
&=&\prod_{i=1}^I \prod
_{j=1}^J \prod_{k=1}^K
P(\mu_{jk}|\bolds\theta)P(y_{ij}, Z1_{ij},Z2_{ij}|
\mu_{jk},\bolds\theta)^{\mu_{jk}}.
\end{eqnarray*}
To estimate the parameters and infer the cluster assignments $\bolds{\mu
}$, we implement the following EM algorithm for a given $K$:
\begin{longlist}
\item[E Step:] This step calculates the expectation of the log
likelihood of parameter $\bolds{\theta}$ conditional on the observed
data and $\bolds{\theta}^{(t)}$ inferred at iteration $t$,
\begin{eqnarray*}
E\bigl(\mu_{jk}|\mathbf{Y}, \bolds{\theta}^{(t)}\bigr)&=&P\bigl\{\mu
_{jk}=1| \mathbf{Y},
\bolds{\theta}^{(t)}\bigr\}
\\[-2pt]
&=&\frac{\pi_k^{(t)}[\prod_{i=1}^I P(y_{ij}, Z1_{ij}, Z2_{ij}|\bolds
{\theta}^{(t)}, \mu_{jk}=1)]}{ \sum_{k=1}^K \pi_k^{(t)}[\prod_{i=1}^I
P(y_{ij}, Z1_{ij}, Z2_{ij}|\bolds{\theta}^{(t)}, \mu_{jk}=1)]}.
\end{eqnarray*}
\item[M Step:] This step updates $\bolds{\theta}$ by $\bolds{\theta
}^{(t+1)}$ that maximizes $Q(\bolds{\theta}|\bolds{\theta}^{(t)})$,
that is,
%
\begin{equation}
\bolds{\theta}^{(t+1)}= \mathop{\arg\max}\limits
_{\bolds{\theta}} Q\bigl(
\bolds{\theta}|\bolds{\theta}^{(t)}\bigr).
\label{maxobjective}
\end{equation}
\end{longlist}
After calculations, we have\vspace*{-5pt}
\[
\pi_k^{(t+1)}=\frac{1}{J}\sum
_{j=1}^JE(\mu_{jk}).
\]
The computation details on the EM algorithm are left in Appendix~\ref
{EM}. Note that the number of~CpG sites in practice can be large, as to
be seen in our real data application. Estimating the scale parameters
$\bolds{\alpha}$ and $ \bolds{\beta}$ in the Beta distribution using
the standard EM algorithm will not be computationally efficient. To
solve this problem, we estimate the scale parameters for a given~CpG
site using the maximum likelihood estimators, based on the sample means
and variances of DNA methylation of each site across all subjects and
the relation between the mean, variance and scale parameters in Beta
distributions. These estimates are then used in the subsequent EM
process to infer $\bolds{\gamma}$ and cluster assignments.
Consequently, we denote our EM algorithm as an empirical EM algorithm.

For parameter $\bolds{\gamma}$, closed-form solutions are not
available. Instead we apply the quasi-Newton method to numerically
maximize \eqref{maxobjective}. The empirical EM algorithm stops when
the increase of the log likelihood from the current iteration to the
next is less than a threshold, for instance, $10^{-7}$.

\subsection{Clustering based on subset sampling}\label{overlap}
In epigenome-wide studies, the data often contain a large number of~CpG
sites, which makes the computation of clustering all~CpG sites at a
time intractable. In this case, we propose to repeat the EM algorithm
for a certain number of subsets of~CpG sites randomly chosen from the
complete data, each with size of $S$. The number of random subsets is
determined such that $100(1-\frac{S}{J})^m\%\leq\eta\%$, where $S$ is
the number of~CpG sites in the subset, $J$ is the total number of~CpG
sites, and $m$ is the number of subsets needed. This is to ensure at
most we miss $\eta\%$ of the~CpG sites after taking $m$ subsets. The
final clusters are determined by a second stage clustering applied to
the inferred $\bolds{\gamma}$ from each subset. Because of the use of
random subsets, it may happen that one~CpG site is included in multiple
clusters. For situations like this, individual likelihood will be
calculated to determine which cluster this~CpG site is more likely to
belong to.

\subsection{Determining the number of clusters $K$}\label{numclust}
To determine the number of clusters, we propose to use the Bayesian
information criterion (BIC) \citep{Schwarz19786} to choose $K$. The BIC
is defined as
%
\begin{equation}
\mathrm{BIC}_K=-2l+(6J+4K-1)\log(3\times I\times J), \label{bic}
\end{equation}
where $l$ denotes the log-likihood $\log(L(\bolds{\theta}|\mathbf{Y}))$,
$6J+4K-1$ is the total number of free parameters to be estimated and $3
\times I \times J$ is the number of observations. To determine the
number of clusters, we propose to use the idea in the screen plot of
BICs versus the corresponding numbers of clusters. Screen plots are
often used in principal component analysis to determine the number of
components, where a sharp decrease in eigenvalues indicates less
importance of the rest of the components. Analogously, in our
application of screen plots, a sharp decrease in BIC indicates that
large numbers of clusters are less preferred. The method discussed in
this section is programmed in R and available to researchers of interest.

\section{Numerical analysis}\label{Nume}
\subsection{Simulations}\label{simu} The proposed method is
demonstrated and evaluated by use of 100 Monte Carlo (MC) replicates.
Each MC replicate represents DNA methylation of 2000~CpG sites from 60
triads generated from Beta distributions. The scale parameters in the
Beta distribution are randomly selected for each~CpG site, which
potentially results in unique patterns of data at each~CpG site. Sixty
triads are used, as it is close to the number of triads in our real
data. These 2000~CpG sites were assigned into 4 clusters with
coefficients $\bolds{\gamma}_1 =(-4.2, 0, 1.3)$, $\bolds{\gamma}_2
=(-0.7, 1.9, 0)$, $\bolds{\gamma}_3 =(-2.3, 0, 0)$, $\bolds{\gamma}_4
=(1.4, -1.5, -0.6)$, with the first cluster representing~CpG sites
fully paternal transmission, the second fully maternal transmission,
the third cluster representing a situation that DNA methylation of
these sites are not transmitted and the fourth composed of~CpG sites
with DNA methylation transmission dominated by mother. Each cluster is
of size 500, that is, 500~CpG sites.

\begin{figure}[t]

\includegraphics{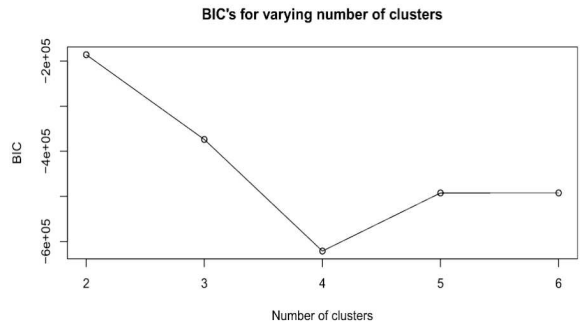}

\caption{BIC values for a varying number of clusters in the simulated
data with 2000~CpG sites. The BIC curve drops down greatly at the
beginning and achieves its minimum at 4.}\label{BIC-smallCpG}
\end{figure}

We then apply the proposed method to perform the cluster analyses for
each MC replicate. To summarize our results, for each MC replicate, we
record the number of clusters identified and calculate the sensitivity
and specificity of the clustering, based on which we calculate the mean
specificity and sensitivity along with their standard deviations. As
noted earlier, the number of clusters is determined by use of the
screen plot of BICs with each BIC corresponding to a specific number of
clusters. As an illustration, Figure~\ref{BIC-smallCpG} shows the
pattern of BIC from one MC replicate, from which we infer 4 clusters.

\begin{table}[b]
\tabcolsep=0pt
\tablewidth=252pt
\caption{The occurrence frequency of each cluster number over 100 repetitions}\label{occrrence}
\begin{tabular*}{\tablewidth}{@{\extracolsep{\fill}}@{}lccccc@{}}
\hline
\textbf{Number of clusters ($\bolds{K}$)} &\textbf{2}& \textbf{3} & \textbf{4} & \textbf{5} & \textbf{6}\\
\hline
Frequency& 0& 1 & 92& 4 & 3 \\
\hline
\end{tabular*}
\end{table}

The uncertainty on the number of clusters is given in Table~\ref
{occrrence} as the occurrence frequency for each cluster number over
100 MC replicates. The median of the number of clusters is 4 with a
$95\%$ empirical interval of $(3, 6)$. After a closer investigation on
the situations where 5 or 6 clusters were inferred, we found a minimal
decrease in BIC from 4 clusters to 5 or 6 clusters compared to the
difference between 3 and 4 clusters. This implies that BIC reaches a
plateau at 4 clusters, and thus 4 clusters were selected following the
concept of the screen plot. The sensitivity and specificity for
identifying each cluster over these 100 replicates are summarized in
Table~\ref{roc-smallCpG}. In general, sensitivity and specificity are
high for all 4 clusters with small variations.

A question may be raised regarding the performance of the proposed
clustering method in comparison with the existing methods noted in
Section~\ref{Intro}. Since the proposed method performs clustering
analyses directly on the strength of inheritance at the population
level, it is expected to be more sensitive compared to the existing
methods. For the purpose of demonstration, we use the $K$-means approach
as an example. For each of the 100 MC replicates, we first calculate
sample mean methylations at each~CpG site for father, mother and their
offspring, respectively, and then apply the $K$-means method to these
mean methylations. The $K$-means approach can be implemented using the R
function $K$-means. Summary statistics on sensitivity and specificity
across the 100 MC replicates is presented in Table~\ref{roc-smallCpG}.

\begin{table}[t]
\tabcolsep=0pt
\caption{Average sensitivity and specificity across 100 repetitions for
each inferred cluster. PROP denotes the proposed Beta regression
clustering method. There are 500~CpG sites in each cluster}\label{roc-smallCpG}
\begin{tabular*}{\tablewidth}{@{\extracolsep{\fill}}@{}lcccccc@{}}
\hline
\textbf{Cluster index} && & \textbf{1} & \textbf{2} & \textbf{3} & \textbf{4}\\
\hline
Sensitivity& PROP & Mean & 0.9600 & 1.0000 & 0.9554 & 0.9749 \\
& & SD & 0.1969 & 0.0000 & 0.2014 & 0.1488\\[3pt]
& $K$-means & Mean & 0.9137 & 0.9659 & 0.6300 & 0.9443 \\
& & SD & 0.1852 & 0.1116 & 0.4852 & 0.1522
\\[6pt]
Specificity & PROP & Mean& 0.9561 & 0.9901 & 0.9620 & 0.9753 \\
& & SD & 0.1989 & 0.0565 & 0.1767 & 0.1185\\[3pt]
& $K$-means & Mean & 0.9712 & 0.9886 & 0.8767 & 0.9814 \\
& & SD & 0.0617 & 0.0372 & 0.1617 & 0.0507\\
\hline
\end{tabular*}  
\end{table}

Overall, both the proposed method and $K$-means work well in terms of
sensitivity and specificity. Higher sensitivity of the proposed method
across all 4 clusters indicates it has higher power in identifying~CpG
sites correctly. For clusters 2~and~3, both sensitivity and specificity
from the proposed method are higher on average than those from the
$K$-means method, especially for cluster 3 ($50\%$ higher in sensitivity
and $10\%$ higher in specificity). For clusters 1 and 4, the proposed
method shows higher sensitivity and slightly lower specificity on
average than the $K$-means method. Overall, our method performs better
than the $K$-means approach. This is likely due to the fact that the
$K$-means approach looks for similarity between~CpG sites with respect to
DNA methylation in triads instead of examining transmission patterns as
in the proposed method.~CpG sites showing similar patterns in triad
DNA-M patterns may have different transmission patterns. Recall our
ultimate goal is to infer the clusters as well as the strength of
inheritance from mother and from father (i.e., estimating $\bolds\gamma
$). Should the $K$-means approach be used to cluster the~CpG sites, we
had to go through an additional step to infer the strength of
inheritance for each cluster. In general, we thus expect the proposed
method to perform better and to be more efficient.

Another possible concern is related to the robustness of the method
with respect to the Beta distribution assumption on DNA methylation
measures. To demonstrate this, we simulate 100 MC replicates such that
DNA methylation is generated from truncated normal distributions within
interval $(0,1)$. The mean methylation for parents is set at 0.5, and
for offspring it is calculated through (\ref{inheritance}), where we
use the same coefficients as in previous simulations. The variance in
the truncated normal is set at 0.25 for all triads. With this type of
distribution that is different from Beta distribution, high sensitivity
and specificity are still observed for all clusters (results not
shown). This set of simulations provides evidence that data
distributions may not substantially affect the clustering result.

\subsection*{Further assessment of the method}
The above simulations are performed on the~CpG sites which are evenly
distributed in 4 clusters, that is, each cluster has equal size of 500
CpG sites and all the~CpG sites are assumed to be independent. What
will happen if the clusters are uneven, the~CpG sites are correlated in
DNA methylation, or there are more inheritance patterns other than
four? To evaluate the proposed method comprehensively, four additional
simulation scenarios are considered, denoted by S1, S2, S3, S4,
respectively. For convenience, we denote the previous simulation
scenario as scenario S0:
\begin{enumerate}[S1]
\item[S1] Unbalanced cluster size in terms of the number of~CpG sites,
a more realistic scenario in practice. We revise scenario S0 by taking
unbalanced cluster sizes such that the numbers of~CpG sites are 500,
600, 850 and 50 for clusters 1 to 4, respectively. Other settings are
the same as in scenario S0 (the same note in the following scenarios).
In the subsequent scenarios, all clusters are unbalanced.
\item[S2] Correlated~CpG sites in DNA methylation. We revise scenario
S1 by generating~CpG sites in clusters 1 and 2 such that correlation of
DNA methylation in neighboring~CpG sites is~0.90. Note that in this
scenario, the number of~CpGs in each cluster is the same as in S1, that
is, not balanced. This setting will assess the robustness of the method
with respect to correlated~CpGs as well as unbalanced clusters.
\item[S3] More varieties in parental effects. Instead of the four
parental effects given in S0, we added another effect with coefficients
$\bolds{\gamma}_5=(-3,2,2)$, that is, parental transmission is evenly
distributed between mother and father. The numbers of~CpG sites in
clusters 1 to 5 are 500, 600, 450, 50 and 400, respectively.
\item[S4] Large number of~CpG sites. Scenario S0 considers 2000~CpG
sites. We expand the number of~CpG sites to 10,000 and include 2500,
3000, 4250, and 250~CpGs in clusters 1 to 4.
\end{enumerate}
All these additional scenarios are designed in order to evaluate the
sensitivity and robustness of the proposed method. For each scenario,
100 MC replicates are generated and each is with sample size of 60
subjects. We use the same statistics as for S0 to summarize the
findings, and we also compare with the findings from $K$-means.

\begin{table}[t]
\tabcolsep=0pt
\caption{Average sensitivity and specificity across 100 repetitions for
each inferred cluster for scenario \textup{S1} where there are 500, 600, 850, 50
CpG sites in clusters 1, 2, 3 and 4, respectively}\label{setting1}
\begin{tabular*}{\tablewidth}{@{\extracolsep{\fill}}@{}lcccd{1.5}d{1.5}c@{}}
\hline
\textbf{Cluster index} && & \textbf{1} & \multicolumn{1}{c}{\textbf{2}} & \multicolumn{1}{c}{\textbf{3}} & \textbf{4}\\
\hline
Sensitivity& PROP & Mean & 0.9625 & 0.9691 & 0.9714 & 0.9511 \\
& & SD & 0.1708 & 0.1385 & 0.1390 & 0.1958\\[3pt]
& $K$-means & Mean & 0.8494 & 0.9557 & 0.6736 & 0.3224 \\
& & SD & 0.2345 & 0.1318 & 0.3591 & 0.4498
\\[6pt]
Specificity & PROP & Mean& 0.9686 & 0.98838 & 0.98843 & 0.9590 \\
& & SD & 0.1713 & 0.10004 & 0.09998 & 0.1968 \\[3pt]
& $K$-means & Mean & 0.9498 & 0.9810 & 0.7587& 0.9826 \\
& & SD & 0.0782 & 0.0565 & 0.2654 & 0.0115\\
\hline
\end{tabular*}  
\end{table}

\begin{table}[b]
\tabcolsep=0pt
\caption{Average sensitivity and specificity across 100 repetitions for
each inferred cluster for scenario \textup{S2} where there are 500, 600, 850, 50
CpG sites in clusters 1, 2, 3 and 4, respectively. Any two consecutive
CpG sites in clusters 1 and 2 are highly correlated}\label{setting2}
\begin{tabular*}{\tablewidth}{@{\extracolsep{\fill}}@{}lcccccc@{}}
\hline
\textbf{Cluster index} && & \textbf{1} & \textbf{2} & \textbf{3} & \textbf{4}\\
\hline
Sensitivity& PROP & Mean & 0.9051 & 0.9124 & 0.8498 & 0.8667 \\
& & SD & 0.2688 & 0.2370 & 0.3334 & 0.3103\\[3pt]
& $K$-means & Mean & 0.9327 & 0.8919 & 0.7924 & 0.5230 \\
& & SD & 0.1510 & 0.1558 & 0.3187 & 0.4759
\\[6pt]
Specificity & PROP & Mean& 0.9580 & 0.9978 & 0.9085 & 0.9862 \\
& & SD & 0.0659 & 0.0038 & 0.1626 & 0.0270\\[3pt]
& $K$-means & Mean & 0.9776 & 0.9537 & 0.8466& 0.9878 \\
& & SD & 0.0503 & 0.0668 & 0.2356 & 0.0122\\
\hline
\end{tabular*}
\end{table}

\begin{table}[t]
\tabcolsep=0pt
\caption{Average sensitivity and specificity across 100 repetitions for
each inferred cluster for scenario \textup{S3} where there are 500, 600, 450,
50, 400~CpG sites in clusters 1, 2, 3, 4 and 5, respectively. Both
clusters 4 and 5 have nonzero coefficients for parents}\label{setting3}
\begin{tabular*}{\tablewidth}{@{\extracolsep{\fill}}@{}lccccccc@{}}
\hline
\textbf{Cluster index} && & \textbf{1} & \textbf{2} & \textbf{3} & \textbf{4} &\textbf{5} \\
\hline
Sensitivity& PROP & Mean & 0.8391 & 0.8248 & 0.8798 & 0.8601 & 0.8372 \\
& & SD & 0.3219 & 0.3286 & 0.2640 & 0.2955 & 0.3115\\[3pt]
& $K$-means & Mean & 0.9030 & 0.9065 & 0.9371 & 0.4968 & 0.5390 \\
& & SD & 0.1723 & 0.1886 & 0.1782 & 0.4780 & 0.2811
\\[6pt]
Specificity & PROP & Mean& 0.9311 & 0.8784 & 0.9090 & 0.9539 & 0.9095 \\
& & SD & 0.1258 & 0.2906 & 0.2027 & 0.1455 & 0.2336\\[3pt]
& $K$-means & Mean & 0.9677 & 0.9599 & 0.9817 & 0.9871 & 0.8848 \\
& & SD & 0.0574 & 0.0808 & 0.0517 & 0.0123 & 0.0703\\
\hline
\end{tabular*} 
\end{table}

\begin{table}[b]
\tabcolsep=0pt
\caption{Average sensitivity and specificity across 100 repetitions for
each inferred cluster for scenario S4 where there are 2500, 3000, 4250,
250~CpG sites in clusters 1, 2, 3 and 4, respectively, with the total
number of 10,000~CpG sites. The number of~CpG sites in each cluster is
5 times the number of~CpG sites in scenario \textup{S1}. Note here PROP uses
five digits to avoid the confusion of equal mean and SD of specificity
with four digits between clusters 1 and 3, 2 and 4}\label{setting4}
\begin{tabular*}{\tablewidth}{@{\extracolsep{\fill}}@{}lcccccc@{}}
\hline
\textbf{Cluster index} && & \textbf{1} & \textbf{2} & \textbf{3} & \textbf{4}\\
\hline
Sensitivity& PROP & Mean & 0.95872 & 0.96778 & 0.94827 & 0.96819 \\
& & SD & 0.19671 & 0.17109 & 0.21824 & 0.17116\\[3pt]
& $K$-means & Mean & 0.8617\phantom{0} & 0.9618\phantom{0} & 0.7535\phantom{0} & 0.3870\phantom{0} \\
& & SD & 0.2236\phantom{0} & 0.1244\phantom{0} & 0.3189\phantom{0} & 0.4831\phantom{0}
\\[6pt]
Specificity & PROP & Mean& 0.95881 & 0.96811 & 0.95880 & 0.96847 \\
& & SD & 0.19682 & 0.17129 & 0.19679 & 0.17132\\[3pt]
& $K$-means & Mean & 0.9539\phantom{0} & 0.9836\phantom{0} & 0.8178\phantom{0}& 0.9841\phantom{0} \\
& & SD & 0.0745\phantom{0} & 0.0533\phantom{0} & 0.2358\phantom{0} & 0.0124\phantom{0}\\
\hline
\end{tabular*}
\end{table}

The results for these scenarios S1--S4 are summarized in Tables~\ref
{setting1}, \ref{setting2}, \ref{setting3} and \ref{setting4}. For
scenario S1 (unbalanced cluster sizes), high sensitivities and
specificities are observed across all clusters with small variations
over 100 MC replicates (Table~\ref{setting1}). As a comparison, when
the number of~CpGs in a cluster is small (i.e., 50~CpG sites in cluster
4), the $K$-means method produces much lower sensitivity. For scenario S2
(correlation in DNA methylation in clusters 1 and 2 is~0.90), the
findings from the proposed method is not greatly influenced by the high
correlations in DNA methylation between neighboring~CpG sites. High
sensitivities and specificities on average are still present and in
general better than the results from the $K$-means approach. The large
impact on sensitivity of the $K$-means method when the number of~CpGs is
small is still present (Table~\ref{setting2}). Turning to scenario S3
(more varieties in parental effects), reasonably high sensitivities and
specificities are present (Table~\ref{setting3}). The $K$-means method
gives slightly better statistics for clusters 1 to 3, but for clusters
4 and 5, the results from the proposed methods are much better. In all
these three scenarios, we observed that the proposed approach results
in consistently high sensitivities and specificities, even for clusters
with a small number of~CpG sites. In scenarios S0, S1, S2, S3, 2000~CpG
sites are considered. In reality, the number of~CpG sites can be much
larger. This is the motivation of scenario S4 proposed above. As seen
in Table~\ref{setting4}, the proposed method is not influenced by the
number of~CpG sites and performs well with high sensitivities and
specificities. Furthermore, it outperforms the $K$-means approach,
especially for cluster 4 (with 250~CpG sites).

Although in general the average assessment statistics (sensitivity and
specificity) from the proposed method are better than those from the
$K$-means approach, we noted that, across the MC replicates, the
variations of these statistics, especially the variations of
specificity, are often smaller for the $K$-means approach (except for the
situations that the $K$-means performs substantially inferior to the
proposed approach). This is likely due to the stronger requirement of
the proposed method, that is, clustering based on associations instead
of means only as in the $K$-means approach. Nevertheless, these
simulations demonstrate that the proposed approach has the ability to
correctly cluster~CpG sites based on parental inheritance of DNA
methylation. It is on average insensitive to the number of~CpGs in each
cluster, robust with respect to the correlation in DNA methylation
between neighboring~CpGs, and is able to handle a variety of parental
effects and a large number of~CpG sites.

\subsection{Real data analysis}
We applied the proposed clustering method to DNA methylation data of
the $F_1$ and $F_2$ generation on the Isle of Wight cohort. This birth
cohort was established in 1989--1990 aiming to study the natural history
of allergic disease \citep{SyedHasanArshad199290a}. In this study, 41
triads (mother, father and child) are included with mother or father
from the 1989--1990 birth cohort. For parents, methylation was
determined in DNA extracted from blood samples (peripheral blood
leucocytes) collected at the time of pregnancy; and for children, DNA
methylation was determined in DNA extracted from cord blood.

Genome-wide DNA methylation was assessed using a technology similar to
genotype identification.\footnote{Illumina Infinium HumanMethylation450
BeadChip (Illumina, Inc., San Diego, CA, USA).} The genome-wide DNA
methylation data covers over 484,000~CpG sites associated with
approximately 24,000 genes. The methylation level for each queried~CpG
is presented as beta values. They represent the proportions of
intensity of methylated ($M$) over the sum of methylated and
unmethylated ($U$) sites, $\operatorname{beta}=M/[c+M+U]$ with constant $c$ introduced
for the situation of too small $M+U$. The value of $c$ was determined
by the company generating the DNA methylation data and its value is
usually taken as $c=100$.

The methylation data were preprocessed using the Bioconductor IMA
package and the ComBat function in R for initial quality control to
remove unreliable~CpG sites, correct for probe types, remove background
noise and correct for batch effect \citep{DanWang201228,Johnson20078}.
After preprocessing and batch effect removal, 308,000 sites were
retained for the next step of screening. Since our goal is to identify
transmission patterns, we use the screening to exclude~CpG sites with
weak correlations in DNA methylation between parent and child. A~CpG
site will be excluded from further consideration if the mother-child or
father-child correlation in DNA methylation is $<$0.5. This screening
process resulted in 4063~CpG sites on autosomes (nonsex chromosomes)
for all 41 triads, which are included in the cluster analysis. Note
that the screening may cause missingness of~CpG sites such that DNA
methylation is transmitted at the population level at those~CpG sites.
Our plan was to focus on~CpG sites of which DNA methylation between
offspring and parents showed at least moderate correlations. Under this
context,~CpG sites showing different transmission patterns at the
population level may be of particular interest; correlations address
the connection between offspring and parents at the individual level,
while transmission patterns inform at the population level how DNA
methylation in offspring is controlled by parents' DNA methylation.~CpG
sites showing weak or no connections between offspring and parents may
not be of great interest.

\begin{figure}[b]

\includegraphics{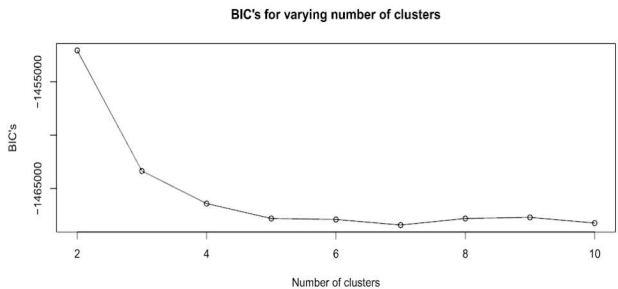}

\caption{BIC values for a varying number of clusters for the IOW real
data with 4063~CpG sites and 41 triads. We extend the searching range
of the number of clusters compared to the simulation study with the aim
to get a safe result. The BIC curve decreases sharply, then reaches the
minimum when $K=7$ but 6 clusters are obtained.}\label{BIC-realdata}
\end{figure}

The empirical EM algorithm discussed in Section~\ref{emsection} is
applied to estimate the parameters and assign~CpG sites to different
clusters. BIC defined in Section~\ref{bic} is used to estimate the
number of clusters. BICs with respect to a different number of clusters
are displayed in Figure~\ref{BIC-realdata}. The BIC achieves its
minimum at $K=7$. A slight difference in BIC between $K=7$ and $K=6$ is
observed. In addition, with $K=7$, there is a null cluster that has no
CpG sites included. All these plus the implementation of screen plot
indicate that 6 clusters are preferred. The estimated coefficients
explaining transmission strength, the standard errors calculated using
100 bootstrap samples, and the numbers of~CpG sites included in each
cluster are summarized in Table~\ref{dissum-realdata}. All the standard
errors are small compared to the corresponding estimated coefficients
implying high confidence in the estimates.

\begin{table}
\tabcolsep=0pt
\caption{Coefficient estimate and distribution summary of 4063~CpG
sites in each cluster. SE denotes the standard errors using the
Bootstrap method over 100 repetitions}\label{dissum-realdata}
\begin{tabular*}{\tablewidth}{@{\extracolsep{\fill}}@{}lcccc@{}}
\hline
\textbf{Cluster}      &                                              & $\bolds{\widehat{\gamma}_{1k}}$ \textbf{(SE)} &  $\bolds{\widehat{\gamma}_{2k}}$ \textbf{(SE)} & \textbf{No.}\\
\textbf{index}        & $\bolds{\widehat{\gamma}_{0k}}$ \textbf{(SE)}& \textbf{maternal}                             & \textbf{paternal} & \textbf{CpG}\\
\textbf{($\bolds{k}$)}&                                              &\textbf{transmission}                          & \textbf{transmission} & \textbf{sites}\\
\hline
1& \phantom{$-$}0.2695 (0.0119) & 0.5704 (0.0093) & 0.4638 (0.0091) & \phantom{0}349 \\
2& \phantom{$-$}0.7019 (0.0187) & 0.2151 (0.0174) & 0.8547 (0.0181) & \phantom{00}53\\
3& \phantom{$-$}1.1761 (0.0341) & 0.6727 (0.0203) & 0.4763 (0.0211) & \phantom{00}14\\
4 & $-$0.2357 (0.0124) & 0.5415 (0.0143) & 0.5236 (0.0141) & 2182\\
5 & \phantom{$-$}0.4783 (0.0269) & 0.5265 (0.0299) & 0.5106 (0.0281) & \phantom{0}118\\
6 & \phantom{$-$}0.0536 (0.0578) & 0.6414 (0.0626) & 0.3808 (0.0507) & 1347\\
\hline
\end{tabular*}
\end{table}

\begin{figure}[t]

\includegraphics{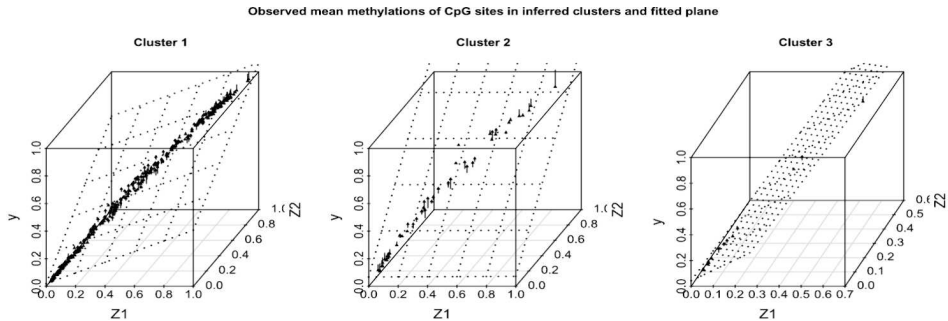}

\caption{Circle points are observed methylations of~CpG sites. Dashed
lines indicate the fitted plane using coefficient estimates in
Table~\protect\ref{dissum-realdata}. At each circle point, residuals
between fitted plane and observed mean methylations are displayed by
vertical solid lines.}\label{realplot}
\end{figure}

Recall that parameter $\gamma_{1k}$ represents the strength of maternal
transmission and $\gamma_{2k}$ the strength of paternal transmission.
Among the 6 identified clusters, cluster 2 containing 53~CpGs (their
locations and corresponding genes are given in Appendix~\ref{appen})
was predominantly paternal-transmitted, as indicated by the larger
estimate of $\gamma_{2k}$; with maternal DNA methylation held constant,
10\% increase in paternal DNA methylation in the population will result
in a 0.08547 increase in the offspring population DNA methylation, but
it will be only a 0.02151 increase should maternal DNA methylation in
the population increase by one unit. The intercept $\gamma_{0k}$ will
be practically meaningful only when neither of the parents transmit
their DNA methylation to their offspring, in which case it represents
the average DNA methylation of a child. In this case, it is likely that
the mother had minimal contribution to offspring DNA methylation.
Following the same way of interpreting the coefficients, clusters 3 and
6 (together containing 1361~CpGs) were mainly maternal-transmitted.
The remaining clusters showed a comparable transmission pattern between
mothers and fathers. To give a general impression of various mean
patterns for~CpG sites in different clusters, we plotted the mean
methylations of each~CpG site in clusters 1, 2 and 3, together with the
plane that the fitted line is in (Figure~\ref{realplot}), where
different patterns in different clusters are shown.

We further examined the biological functions associated with~CpG
clusters exhibiting paternal or maternal bias in transmission. Both
maternally and paternally transmitted clusters were significantly
enriched for genes that contain genetic polymorphisms and are regulated
by alternative splicing ($p < 0.05$ after Benjamini--Hochberg
correction [\citet{YoavBenjamini199557}]). The most significant
membrane-related term in the maternally transmitted clusters was
``glycoprotein'' (25 genes, 1.9-fold enriched, $p = 0.063$ after
Benjamini--Hochberg correction). The fatty acid composition of the
plasma membrane is associated with allergic disease risk \citep
{I.Romieu200737}; also, the enrichment of membrane-related terms is
concordant with an effect on allergy and immunity, as the cellular
membrane holds many immune-related proteins on the cell surface.

The maternally transmitted clusters also include genes functionally
linked to allergic disease, such as \textit{HLA-B}, which encodes an MHC class I
peptide involved in antigen presentation and has a well-known
association with atopy, located in the HLA region which itself has been
associated with asthma in multiple GWASs \citep
{GabrielleALockett20135}. Comparably, the paternally transmitted
cluster as well contains genes known to be functionally linked to
allergy and immunity, including \textit{IL18BP}, which encodes a binding protein
for IL18 in the Th1 immunity pathway; \textit{TLR4} which encodes a receptor on
the surface of immune cells for detecting gram negative bacteria; and
\textit{BAT3} which is associated with \textit{HLA-B}, a gene extensively linked to
allergic disease as described above.

\subsection*{Findings from the subset sampling approach}
We also used the subset sampling approach described in Section~\ref
{overlap} on the 41 triads data set. We set $S=2000, m=15$ such that
every~CpG site is chosen for clustering at least once. We run the
empirical EM algorithm with $K$ varying. The subset sampling approach
identified 3 clusters instead of 6. However, the corresponding BIC was
larger than that when~CpG sites were used all at once. Among these 3
clusters, there are 325~CpG sites in one cluster with $\widehat{\gamma
}_{2k}$ higher than $\widehat{\gamma}_{1k}$, indicating these 325~CpG
sites may belong to a paternally transmitted cluster. It was found that
these 325~CpG sites contain all 53~CpG sites identified via clustering
all~CpG sites at the same time. Furthermore, for~CpG sites that are
equally transmitted or maternally transmitted, large overlaps were
observed as well. All these provide evidence that these two approaches
can reach similar conclusions. However, if the number of~CpG sites is
not large, we recommend using all~CpG sites to perform the analysis, as
it is expected to give a better fit. The subset-based sampling approach
is recommended if the number of~CpG sites is extremely large.

\subsection*{Further investigations on inheritance}

\begin{figure}

\includegraphics{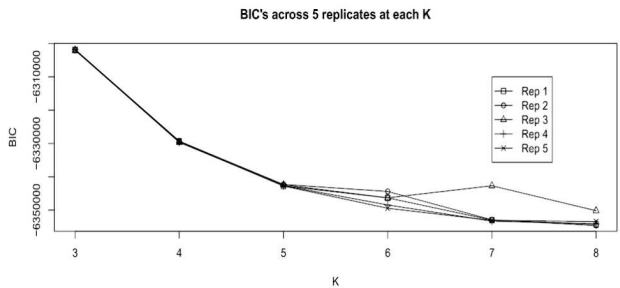}

\caption{BIC vs. number of clusters across five randomly chosen seeds
indicated by Rep $1, 2, \ldots, 5$.}\label{biccutoff0.4}
\end{figure}
The above findings are based on candidate~CpG sites obtained by
implementing a cutoff of 0.5 in correlations. We further relaxed the
correlation cutoff to 0.4, which resulted in 14,845 candidate~CpG
sites. The proposed clustering algorithm is implemented on this larger
data set (14,485~CpGs on 41 triads). The same number of clusters (6
clusters) are determined based on the screen plot of the BICs. To
eliminate the possibility due to random sampling, we implemented
different seeds in random number generators. The same number of
clusters was inferred with different seeds (Figure~\ref{biccutoff0.4}).
Among the 6 clusters, the cluster patterns in general are in agreement
with those when the cutoff is 0.5, except that one cluster (with 62~CpG
sites) was identified such that mother's effect was close to zero
(coefficient was 0.0097), that is, no maternal inheritance. These 62
CpG sites were not identified when the cutoff was 0.5. Furthermore, all
the 53~CpG sites, which were inferred previously as predominately
paternal-transmitted~CpG sites, were still grouped into the same
cluster along with the additional 300~CpG sites. This finding is
expected due to the expanded set of candidate~CpGs. However, in these
15K~CpG sites, we did not identify~CpG sites that are completely
untransmitted at the population level.

The above findings motivated us to further investigate the proposed\break
method in its ability to identify nontransmitted~CpG methylation. We
considered the following two scenarios, and, for each scenario, we
generated five data sets:
\begin{longlist}[2.]
\item[1.] Simulate nontransmitted data based on real data. We randomly
extracted 2000~CpG sites from the real data with cutoff 0.5. DNA
methylation of all the triads are as in the real data except for the
last 500~CpG sites, of which DNA methylation for the offspring are
generated based on their parents' methylation such that the regression
coefficients are set at zero but intercept is nonzero. These 500~CpG
sites represent nontransmitted~CpGs in DNA methylation.

\item[2.] Select candidate~CpGs such that correlations are $<0.1$. We
considered five data sets with each composed of randomly selected 1000
CpGs that satisfy this requirement (i.e., correlations $<0.1$). Note
that low correlations at the individual level can possibly lead to
(since whether a child has a high DNA methylation at a~CpG site has
nothing to do with his/her parents' DNA methylation at that site), but
is not equivalent to, nontransmission at the population level. For
instance, it is still possible that at the population level, on
average, higher DNA methylation of father and mother at a particular
CpG site results in high DNA methylation in offspring. Because of the
low correlations, offspring's DNA methylation is likely not to be
connected to parents' DNA methylation and, consequently, these types of
CpGs may not be of great interest.
\end{longlist}

In the first scenario, across all the five simulated data sets, a truly
nontransmitted~CpG site was included in the nontransmission cluster
with high probability (ranged from 0.52 to 0.88), which provides
further evidence that the proposed method has the ability to identify
nontransmitted~CpGs. In the second scenario, for each set of randomly
selected 1000~CpGs, we identified a small portion ($1\%$) of~CpGs
showing nontransmission (indicated by regression coefficients close to
zero). This finding supports our expectation noted above. That is, DNA
methylation at~CpG sites showing low correlations at the individual
level may still be transmitted at the population level. Furthermore,
this also implies that at the population level DNA methylation is more
likely to be transmitted from one generation to the next.

Summarizing all the above investigations, we postulate that DNA
methylation at most~CpGs is transmitted equally from the two parents to
the next generation, a much larger number of~CpGs are
maternal-transmission dominated than those from paternal-transmission,
and only at a small number of~CpGs DNA methylation is not transmitted
to the next generation. However, future epigenetic research is
certainly deserved to investigate this postulation, for example, by
applying the method to different independent cohorts and assessing
agreement in identified clusters.

\section{Summary and discussion}\label{SuDi}

In conjunction with genetic factors, epigenetics could allow us to
better explain disease transmission from parents to offspring.~CpG
sites showing maternally or paternally biased transmission are of
particular relevance to allergic disease, given that allergic diseases
are inherited in an asymmetric manner \citep{S.HasanArshad2012130}. It
is therefore a matter of great importance to identify~CpG sites showing
maternally and paternally biased transmission of DNA methylation, as
these may permit transgenerational epigenetic transmission of allergic
disease risk. To this end, we proposed the clustering method built upon
the empirical EM algorithm to cluster~CpG sites based on the
relationship of DNA methylation transmission between parents and their
offspring. Candidate~CpG sites used in the cluster analysis were
obtained from a whole-genome screening process using correlations in
DNA methylation between parents and their offspring.

Although DNA methylation of most~CpG sites was transmitted from father
and mother equally, there were a large number of~CpG sites where
maternal influenced as DNA methylation was stronger. Greater maternal
influence was expected, given the stronger maternal influence of the
intrauterine environment. An interesting finding is the identification
of a small set of~CpG sites where DNA methylation is paternally
transmitted. Paternal transmission could represent \textit{bona fide}
transmission of DNA methylation through the germline. Paternal effects,
which must be transmitted via epigenetics, have been observed
previously \citep
{THEODOREJ.CICERO1991256,M.Ledig199837,DawnM.Bielawski200226,FangHe200628,LillianA.Ouko200933}.
We cannot exclude the possibility that apparent maternal and paternal
transmission at some loci could be a product of shared environmental
factors, though, as the child's methylation was assessed at birth
before it is directly exposed to the shared environment, this effect
should be minimal. Our findings implied DNA methylation at a small
number of~CpGs not transmitted to the next generation, which needs
further investigation.

The methodology proposed in this work is not limited to DNA methylation
data and can be applied to other types of data ranged from 0 to 1, for
instance, proportions of successes. It is possible that the
transmission is nonlinear. In situations like this, splines can be
implemented to approximate the association patterns and the
heterogeneity can be evaluated by, for example, the sum of the
coefficients in the base functions.

\begin{appendix}
\section{EM algorithm}\label{EM}
To estimate the parameters and infer the cluster assignments $\bolds{\mu
}$, we implement the following EM algorithm for a given $K$:
\begin{longlist}
\item[E Step:] The $Q$ function at this step is
\begin{eqnarray*}
\label{objectivefunc} Q\bigl(\bolds{\theta}|\bolds{\theta}^{(t)}\bigr
)&=&E_{\bolds{\mu}|\mathbf{Y}, \bolds{\theta}^{(t)}}\bigl[\log\bigl(P(
\mathbf{Y}, \bolds{\mu}|\bolds{\theta})\bigr)\bigr]
\\
&=&\sum_{j=1}^J \sum
_{k=1}^K E\bigl(\mu_{jk}|\mathbf{Y}, \bolds{\theta}^{(t)}\bigr)\log(\pi_k)
\\
&&{}+\sum_{i=1}^I \sum
_{j=1}^J \sum_{k=1}^K
E\bigl(\mu_{jk}|\mathbf{Y}, \bolds{\theta}^{(t)}\bigr)\log\bigl
[P(y_{ij}, Z1_{ij},
Z2_{ij}|\bolds{\theta})\bigr],
\end{eqnarray*}
where
\begin{eqnarray*}
E\bigl(\mu_{jk}|\mathbf{Y}, \bolds{\theta}^{(t)}\bigr)&=&P\bigl\{\mu
_{jk}=1|\mathbf{Y},
\bolds{\theta}^{(t)}\bigr\}
\\
&=&\frac{p_k^{(t)}[\prod_{i=1}^I P(y_{ij}, Z1_{ij}, Z2_{ij}|\bolds{\theta
}^{(t)}, \mu_{jk}=1)]}{ \sum_{k=1}^K p_k^{(t)}[\prod_{i=1}^I P(y_{ij},
Z1_{ij}, Z2_{ij}|\bolds{\theta}^{(t)}, \mu_{jk}=1)]}.
\end{eqnarray*}
\item[M Step:] By taking derivatives of $Q$ with respect to $\bolds{\pi
}$, we have
\begin{table}\vspace*{30pt}
\tabcolsep=0pt
\caption{Relevant information on 53~CpG sites in cluster 2}\label{clust2}
\begin{tabular*}{\tablewidth}{@{\extracolsep{\fill}}@{}ld{2.0}cc@{}}
\hline
\textbf{IlmnID} & \multicolumn{1}{c}{\textbf{\textit{Chromosome}}} & $\bolds{\mathit{UCSC}_{-}\mathit{RefGene}_{-}\mathit{Name}}$ & $\bolds{\mathit{UCSC}_{-}\mathit{RefGene}_{-}\mathit{Group}}$\\
\hline
cg00463982 & 16 & \textit{IFT140}; \textit{TMEM204} & Body; TSS1500\\
cg00484396 &16 & \textit{NAT15} & TSS1500; 5$'$UTR\\
cg00958560 &4 & \textit{C4orf50}& Body\\
cg01571001 &3&&\\
cg01579765 &21& \textit{HSF2BP} & Body\\
cg01757168 &3&\\
cg03536711 &1& \textit{LOC400804} &Body\\
cg03814093 &4& \textit{KIAA0922}& Body\\
cg04230029 &12& \textit{MED13L} & Body\\
cg04495270 &17& \textit{NT5M}& TSS1500\\
cg04753163 &6& \textit{TRERF1} & 5$'$UTR\\
cg05760053 &14& \textit{KHNYN}; \textit{CBLN3} & TSS1500; TSS200\\
cg05767404 &1& \textit{C1orf150} & Body\\
cg07007382 &6& &\\
cg07251788 &22& \textit{CLTCL1} & TSS1500\\
cg07382132 &10& &\\
cg07696842 &12& \textit{CHST11} & Body\\
cg08281415 &16&&\\
cg09516200 &17& \textit{RPTOR} & Body\\
cg09564361 &8&\\
cg11351709 &8&\\
cg11799593 &12&\\
cg12180191 &8& \textit{ANK1} & Body\\
cg13564459 &14& \textit{PRKCH} & Body\\
cg13640690 &9& \textit{LOC100129066} & Body\\
cg13730105 &9 & \textit{TLR4} & Body\\
cg13795986 &1& \textit{RIT1} & Body\\
cg14194983 &1 & \textit{NPPA} & TSS1500\\
cg14314729 &5&&\\
cg14574489 &3 & \textit{SLC9A9}& Body \\
cg14672994 &17 & \textit{ACSF2} & TSS1500\\
cg14734668 &10&&\\
cg14749573 &4 &&\\
cg15937073 &1 & \textit{HIVEP3}& TSS200\\
cg16385335 &11 & \textit{IL18BP}; \textit{IL18B} & TSS200; 5$'$UTR; TSS1500; 5\\
cg16476991 &13 & \textit{RASA3} & Body\\
\hline
\end{tabular*}\vspace*{30pt}
\end{table}
\begin{eqnarray*}
\frac{\partial Q(\bolds{\theta}|\bolds{\theta}^{(t)})}{\partial\bolds
{\pi}}&=&\biggl(\frac{\partial Q(\bolds{\theta}|\bolds{\theta
}^{(t)})}{\partial \pi_1}, \frac{\partial Q(\bolds{\theta}|\bolds
{\theta}^{(t)})}{\partial \pi_2}, \ldots,
\frac{\partial Q(\bolds{\theta}|\bolds{\theta}^{(t)})}{\partial \pi
_{K-1}}\biggr)
\\
&=&\biggl(\frac{\sum_{j=1}^J E(\mu_{j1})}{\pi_1}-\frac{\sum_{j=1}^J E(\mu
_{jK})}{1-\sum_{k=1}^{K-1}\pi_k},
\\
&&{}\frac{\sum_{j=1}^J E(\mu_{j2})}{\pi_2}-
\frac{\sum_{j=1}^J E(\mu_{jK})}{1-\sum_{k=1}^{K-1}\pi_k},
\\
&&{}\ldots, \frac{\sum_{j=1}^J E(\mu_{jK-1})}{\pi_{K-1}}-\frac{\sum
_{j=1}^J E(\mu_{jK})}{1-\sum_{k=1}^{K-1}\pi_k}\biggr)
\stackrel{\triangle} {=}\mathbf{0}_{(K-1) \times1},
\end{eqnarray*}
which yields $\frac{1}{\pi_1}\sum_{j=1}^J E(\mu_{j1})=\cdots= \frac
{1}{\pi_K}\sum_{j=1}^J E(\mu_{jK})$ and, further,
\[
\pi_k^{(t+1)}=\frac{1}{J}\sum
_{j=1}^JE(\mu_{jk}).
\]
\end{longlist}

\section{Relevant information on~CpG sites in clusters~2}\label{appen}
We put relevant information for all 53~CpG sites in cluster 2 in
Table~\ref{clust2} (DNA methylation transmission is paternally
dominated), including~CpG site ID (IlmnID), corresponding gene names
($\mathit{UCSC}_{-}\mathit{RefGene}_{-}\mathit{Name}$), chromosome number and specific locations
($\mathit{UCSC}_{-}\mathit{RefGene}_{-}\mathit{Group}$). We hope this will be helpful to
interested researchers.
\setcounter{table}{7}
\begin{table}
\tabcolsep=0pt
\caption{(Continued)}
\begin{tabular*}{\tablewidth}{@{\extracolsep{\fill}}@{}ld{2.0}cc@{}}
\hline
\textbf{IlmnID} & \multicolumn{1}{c}{\textbf{\textit{Chromosome}}} & $\bolds{\mathit{UCSC}_{-}\mathit{RefGene}_{-}\mathit{Name}}$ & $\bolds{\mathit{UCSC}_{-}\mathit{RefGene}_{-}\mathit{Group}}$\\
\hline
cg19490001 &2& \textit{ANKRD53} & 3$'$UTR; Body\\
cg19906672 &4 & \textit{TBC1D14} & 5$'$UTR\\
cg20654462 &15& &\\
cg21147708 &6 & \textit{SNRNP48} & 3$'$UTR\\
cg21783847 &1 & \textit{CREG1} & Body\\
cg22156674 &2& & \\
cg22508957 &16& \textit{NAT15} & TSS1500; 5$'$UTR\\
cg23474190 &21 &\\
cg24681208 &14& \textit{REM2} & TSS1500\\
cg25229172 &12 & \textit{AMDHD1}; \textit{CCDC38} & TSS1500; 5$'$UTR\\
cg25314284 &11& & \\
cg25651505 &2 & \textit{VAMP5} & Body\\
cg26804772 &1&& \\
cg26929700 &16 & \textit{ZNF423} & Body\\
cg27014438 &6& \textit{BAT3} & Body\\
cg27113548 &14&&\\
cg27448532 &4 & \textit{ARHGAP10} & Body\\
\hline
\end{tabular*}
\end{table}
\end{appendix}

\section*{Acknowledgments}
The authors sincerely thank the Editor, Associate Editor and referees
for their constructive suggestions and comments  which contributed
substantially to the improvement of the work.\newpage



%

\printaddresses
\end{document}